# On the Derivation of Vector Radiative Transfer Equation for Polarized Radiative Transport in Graded Index Media

J.M. Zhao[a], J.Y. Tan[b], L.H. Liu[a]*

[a] *School of Energy Science and Engineering, Harbin Institute of Technology, 92 West Dazhi Street, Harbin 150001, People's Republic of China*

[b] *School of Auto Engineering, Harbin Institute of Technology at Weihai, 2 West Wenhua Road, Weihai 264209, People's Republic of China*

**Abstract**

Light transport in graded index media follows a curved trajectory determined by the Fermat's principle. Besides the effect of variation of the refractive index on the transport of radiative intensity, the curved ray trajectory will induce geometrical effects on the transport of polarization ellipse. This paper presents a complete derivation of vector radiative transfer equation for polarized radiation transport in absorption, emission and scattering graded index media. The derivation is based on the analysis of the conserved quantities for polarized light transport along curved trajectory and a novel approach. The obtained transfer equation can be considered as a generalization of the classic vector radiative transfer equation that is only valid for uniform refractive index media. Several variant forms of the transport equation are also presented, which include the form for Stokes parameters defined with a fixed reference and the Eulerian forms in the ray coordinate and in several common orthogonal coordinate systems.

______________________________________________________________

*Corresponding author. Tel.: +86-451-86402237; fax: +86-451-86221048.

*E-mail address:* lhliu@hit.edu.cn (L. H. Liu).

**Nomenclature**

| | |
|---|---|
| **b** | Binormal of the ray trajectory |
| $c$ | Speed of light in vacuum, m/s |
| **I** | Vector of Stokes parameters |
| **i**, **j**, **k** | Unit vector of the Cartesian coordinate system |
| $I$ | Radiative intensity, W/(m$^2$sr) |
| $n$ | Refractive index |
| **P** | Scattering matrix |
| $Q$ | Second component of **I**, W/(m$^2$sr) |
| **R** | Matrix defined by Eq. (14) |
| **r** | Spatial coordinates vector |
| $s$ | Ray coordinate, m |
| $\mathbf{s}_1$ | An auxiliary vector, $\mathbf{s}_1 = (-\sin\varphi, \cos\varphi, 0)$ |
| **T** | Transform matrix, defined in Eq. (64) |
| $U$ | Third component of **I**, W/(m$^2$sr) |
| $V$ | Fourth component of **I**, W/(m$^2$sr) |
| $x, y, z$ | Cartesian coordinates, m |
| **Z** | Scattering phase matrix |
| $\chi$ | Ellipticity angle of polarization ellipse |
| $\Phi$ | Scattering phase function |
| $\phi_G$ | Rotation angle |
| $\varphi$ | Azimuthal angle |
| $\eta$ | Direction cosine, $\eta = \sin\theta\sin\varphi$ |
| $\kappa$ | Torsion of ray trajectory, m$^{-1}$ |
| $\kappa_a, \kappa_e, \kappa_s$ | Absorption, extinction and scattering coefficient, m$^{-1}$ |
| $\mathbf{\kappa}_a, \mathbf{\kappa}_e$ | Absorption, extinction matrix, m$^{-1}$ |
| $\mu$ | Direction cosine, $\mu = \sin\theta\cos\varphi$ |
| $\mathbf{v}$ | Principal normal of the ray trajectory |
| $\theta$ | Zenith angle |
| $\mathbf{\Omega}, \mathbf{\Omega}'$ | Unit vector of radiation direction |
| $\Omega$ | Solid angle |
| $\xi$ | Direction cosine, $\xi = \cos\theta$ |
| $\psi$ | Orientation angle of the polarization ellipse |
| $\mathfrak{R}$ | Rotation matrix of Stokes vector |
| $\tilde{\mathfrak{R}}$ | Modified rotation matrix, defined in Eq. (31) |

*Subscripts*

| | |
|---|---|
| $b$ | Black body |
| $G$ | Stokes vector defined with Global reference |
| $LF$ | Defined in left-hand frame |
| 1, 2 | Number index of layers |

## 1. Introduction

Polarization is the one most basic characteristic of light. Theory of polarized radiative transfer has found wide applications in the fields of remote sensing and atmospheric optics [1-3] and has potential applications in some other disciplines, such as biomedical optical tomography [4-7].

Due to the structural characteristics of a material or a possible temperature (density) dependency, the refractive index of a media may be inhomogeneous. In the graded index media, light transport follows a curved trajectory determined by the Fermat's principle [8-10]. The variation of refractive index along the ray trajectory will influence the transport of radiative intensity, which has been studied by many researchers [11-16]. If the polarization of the light beam needs to be considered, it is known in geometrical optics that the curved ray trajectory will induce geometrical effects on the transport of polarization ellipse, i.e. the Rytov's field-vector rotation law [9].

Unfortunately, the well known classic vector radiative transfer equation (VRTE) [17] does not take into account the effect of inhomogeneous refractive index distribution, and is theoretically only valid for uniform refractive index media. Up to now, the radiative transfer equation for polarized light transport in graded index media has not been well established in literatures. There were very few works on the derivation of the equation. To our knowledge, Lau and Waterson [18] were the first to publish a derivation of the radiative transfer equation for polarized light transport in graded index media. Their derivation was based on the eikonal approximation of Maxwell equations and used very complicated manipulations. The derived transport equation has not yet been well verified till its establishment. Furthermore, emission is not considered and only scalar extinction is considered in the derivation, and an explicit formulation of the scattering phase matrix including the effect of curved ray trajectory was not provided.

In this paper, by carefully analyzing the conservation quantities for polarized light transport along the curved trajectory, a complete derivation of the vector radiative transfer equation for polarized radiative transport in graded index media (GVRTE) is presented based on a novel approach. The paper is organized as follows. In Section 2, some fundamental physical aspects on the transport of polarized light are reviewed and discussed, in which the focus is on the conservation quantities upon polarized light transport. In Section 3, the GVRTE is derived for non-participating and participating graded index media in the Lagrangian ray coordinate with Stokes vector defined in the local Frenet-Serret frame and a complete definition of the scattering phase matrix is given. In Section 4, several variant forms of the GVRTE are presented, which include the form for Stokes parameters defined with a fixed reference and the Eulerian forms in ray coordinate and in several common orthogonal coordinate systems (e.g. Cartesian, cylindrical, and spherical coordinate systems).

## 2. Physical aspects of polarized light

### *2.1. Description of the polarized light*

It is well known that a light beam of arbitrary polarization can be described by the Stokes parameters $\mathbf{I} = (I, Q, U, V)$ [17, 19, 20], which are intensity formulation of the polarization state of a light beam and are measurable from experiment. The detailed definition of the Stokes parameters for arbitrary polarized beam is referred to Refs. [17, 19, 20] and will not be repeated here. As for an elliptically polarized beam, the polarization state is graphically depicted into a polarization ellipse, and the four components of the Stokes parameters are defined as

$$\mathbf{I} = \begin{pmatrix} I \\ Q \\ U \\ V \end{pmatrix} = \begin{pmatrix} I \\ I\cos 2\chi \cos 2\psi \\ I\cos 2\chi \sin 2\psi \\ I\sin 2\chi \end{pmatrix} \qquad (1)$$

where $\mathbf{I}$ is the Stokes vector, $I$ is the radiative intensity, $\chi$ is the ellipticity angle whose tangent equals the ratio of the axes of the polarization ellipse and $\psi$ is the orientation angle of the polarization ellipse. The definitions of the variables for the polarization ellipse are shown in Fig. 1 for a beam propagates in the *z* direction.

It is noted that the definitions of the Stokes parameters is not unique. There are basically two different types of definition for the Stokes parameters, one is to define them in a right-hand frame (such as in Ref. [20] and in this paper), and the other is to define them in a left-hand frame (such as Refs. [17, 19]). A simple transformation exists between these two definitions, which are discussed in detail in the Appendix A. The different types of definitions of the Stokes parameters will affect the transformation of Stokes parameters to different reference frame, which will further influence the formulation of scattering phase matrix and will be discussed later in Section 3.2.

One important property of the Stokes parameters is that it follows the addition principle for a mixture of independent beam, namely, the Stokes parameters representing a mixture of beams with different polarization can be obtained by the linear addition of the Stokes parameters of each beam. It is this property that allows the Stokes parameters being taken as the working variable to formulate the transport equation.

Following the addition principle, it is easy to show that a partially polarized light can be considered a mixture of unpolarized light and elliptically polarized light. Furthermore, the unpolarized light itself is equivalent to a mixture of two elliptically polarized light with 'opposite' polarization [17]. Hence an arbitrary polarized light can be considered to be the mixtures of elliptically polarized light. The elliptically polarized light in this meaning forms the basis of the space of polarization state. As a result, to understand the transport characteristics of the elliptically polarized light is the key to derive the transport equation for any polarized light.

*2.2. Conserved quantities upon transport*

The governing equation of transport is in essence a conservation equation, which should accurately describe the conservation laws. The establishment of the classical scalar radiative transfer equation and the classical vector radiative transfer equation are based on the conservation of two physical quantities during radiative transport, 1) radiative energy flux, and 2) the Stokes parameters flux. The energy conservation is true in any circumstance. However, the conservation of Stokes parameters should be taken carefully to derive a governing equation for graded index media. That is because the Stokes parameters are affected by the curved ray trajectory, and to formulate the conservation law must take into account the effect of this kind of geometrical effect.

As for polarized light transport along the curved ray trajectory with considering the effect of the geometrical effect, two basic conservation quantities can be determined, 1) radiative energy flux, and 2) the ellipticity angle $\chi$ of the polarization ellipse [9]. Here, the second quantity indicates the conservation of polarization ellipse, namely, the shape of the polarization ellipse is invariant during transport along curved ray trajectory. It is noted that a combination of conserved quantities can form other conserved quantities.

Though the shape of the polarization ellipse is invariant along the curved ray trajectory, the orientation angle of the polarization ellipse is changed due to the geometrical effect. The geometrical effect on the changing of the orientation angle is given by the Rytov's field-vector rotation law [9]. By using the Frenet-Serret frame along the ray trajectory, the Rytov's law shows

$$\frac{d\psi}{ds} = \kappa \tag{2}$$

where $s$ is the ray coordinate and $\kappa$ is the torsion of the trajectory. The definitions of the polarization ellipse in the Frenet-Serret frame along the curved ray trajectory are depicted in Fig. 2, where $\mathbf{\Omega}$ is tangential vector of the ray trajectory (also the propagation direction of the beam), $\mathbf{v}$ is the principal normal and $\mathbf{b}$ is the binormal. The vectors $\mathbf{v}$-$\mathbf{b}$-$\mathbf{\Omega}$ form a *right-hand* frame corresponding to the *x-y-z* frame defined in Fig. 1.

Strictly speaking, $\psi$ is not a conservative quantity as indicated by the Rytov's law. However, it can be considered a special kind of conservation if the term in the right-hand side is moved to the left-hand side. In this case, the left-hand side as a whole can be considered an invariant.

*2.3. Relation between the energy flux and the transport intensity*

The conservation of energy flux (with unit W/m$^2$) can be further formulated as a conservation of radiative intensity (with unit W/m$^2$sr). One well know relation for light transport in transparent graded index media is that the monochromatic intensity divided by the square of the refractive index is invariant along the curved ray trajectory in non-participating medium, namely

$$\frac{d}{ds}\left[\frac{I}{n^2}\right] = 0 \tag{3}$$

where $n$ is the refractive index. This relation is historically known as the Clausius invariant. It can be considered as a relation that describes the conservation of energy flux in the variable of radiative intensity. The relation [Eq.

(3)] can be simply derived based on the conservation of energy by using the layered approximation (as shown in Fig. 3) and the Snell's law of refraction.

The conservation of energy requires

$$I_1 \cos\theta_1 \mathrm{d}\Omega_1 \mathrm{d}A = I_2 \cos\theta_2 \mathrm{d}\Omega_2 \mathrm{d}A \tag{4}$$

namely,

$$I_1 \cos\theta_1 \sin\theta_1 \mathrm{d}\theta_1 \mathrm{d}\varphi \mathrm{d}A = I_2 \cos\theta_2 \sin\theta_2 \mathrm{d}\theta_2 \mathrm{d}\varphi \mathrm{d}A \tag{5}$$

where the subscript denotes the number index of the layer, $\theta$ is the zenith angle, $\varphi$ is the azimuthal angle, $\mathrm{d}\Omega$ is the differential solid angle, $\mathrm{d}A$ is the differential area. It is noted that the refraction only occurs in the optical plane and hence the $\varphi$ is unaffected. The variation of the zenith angle can be determined by the Snell's law as

$$n_1^2 \mathrm{d}\cos^2(\theta_1) = n_2^2 \mathrm{d}\cos^2(\theta_2) \tag{6}$$

which was substituted into Eq. (5) to give

$$\frac{I_1}{n_1^2} = \frac{I_2}{n_2^2} \tag{7}$$

Here the reflection at the interface is intentionally omitted. Similar relation which uses the Fresnel's law to consider the reflection can also be obtained [21]. Equation (3) can be considered as a limit case (when the refractive index difference between the layer is infinitely small) of Eq. (7).

The simple derivation presented above shows a general rule of formulating the conservation of energy flux to the relation of radiative intensity along curved ray trajectory in non-participating medium. As indicated in Section 2.2, a combination of the conserved quantities is also a conservative quantity in non-participating medium. Hence the flux of the fourth component of the Stokes parameters, namely, the flux of $V = I\sin 2\chi$ is also a conserved quantity. With the above derivation in mind, it immediately yields the conservation law for the intensity $V$ as

$$\frac{\mathrm{d}}{\mathrm{d}s}\left[\frac{V}{n^2}\right] = \frac{\mathrm{d}}{\mathrm{d}s}\left[\frac{I\sin 2\chi}{n^2}\right] = 0 \tag{8}$$

Because the orientation angle of the polarization ellipse is affected by the shape of the ray trajectory, the second component ($Q$) and the third component ($U$) of the Stokes parameters are not conserved quantities in traditional understanding. By using a general approach, namely, to take these quantities as conserved quantities, the conservation relation must take into account the geometrical effect on the orientation angle. Detailed treatment is presented in the following section.

**3. Derivation of the transport equation**

*3.1 Non-participating medium: geometrical conservation*

Non-participating medium is a kind of medium without absorption, emission and scattering. The process of light transport in this kind medium is the simplest. In this case, the conservation law along the curved ray trajectory mainly accounts for the geometrical effect caused by the graded index distribution. As in the classic vector radiative transfer equation for polarized radiative transport in uniform index media, the Stokes parameters vector is taken as the working variable. In the following, the conservation of each component of the Stokes parameters is derived.

Referring the definition of the second component $Q$, it yields

$$\frac{d}{ds}\left[\frac{Q}{n^2}\right] = \frac{d}{ds}\left[\frac{I\cos 2\chi}{n^2}\right]\cos 2\psi + \frac{d}{ds}[\cos 2\psi]\frac{I\cos 2\chi}{n^2} \tag{9}$$

As mentioned in previous section, the combination of conserved quantities is also a conservative quantity. Hence the flux of $I\cos 2\chi$ is a conservative quantity, which formulated in intensity form as discussed in Section 2.3 yields

$$\frac{d}{ds}\left[\frac{I\cos 2\chi}{n^2}\right] = 0 \tag{10}$$

Furthermore, by using the Rytov's law [Eq. (2)], Eq. (9) reduces to

$$\begin{aligned}\frac{\mathrm{d}}{\mathrm{d}s}\left[\frac{Q}{n^2}\right]&=-\sin 2\psi\frac{\mathrm{d}}{\mathrm{d}s}[2\psi]\frac{I\cos 2\chi}{n^2}\\&=-2\kappa\frac{I\cos 2\chi\sin 2\psi}{n^2}=-2\kappa\frac{U}{n^2}\end{aligned} \quad (11)$$

This can be considered a general conservation law for the $Q$. Similarly, for the third component $U$, it yields

$$\begin{aligned}\frac{\mathrm{d}}{\mathrm{d}s}\left[\frac{U}{n^2}\right]&=\frac{I\cos 2\chi}{n^2}\frac{\mathrm{d}}{\mathrm{d}s}[\sin 2\psi]\\&=\frac{I\cos 2\chi\cos 2\psi}{n^2}\frac{\mathrm{d}}{\mathrm{d}s}[2\psi]=2\kappa\frac{Q}{n^2}\end{aligned} \quad (12)$$

Until now, the general conservation law for the Stokes parameters is obtained completely, and are given by Eq. (3), Eq. (8), Eq. (11) and Eq. (12). A combination of these equations into vector form can be written as

$$n^2\frac{\mathrm{d}}{\mathrm{d}s}\left[\frac{\mathbf{I}}{n^2}\right]+\mathbf{R}\,\mathbf{I}=0 \quad (13)$$

where $\mathbf{R}$ is a matrix accounting for the rotation of the polarization ellipse determined by the Rytov's law defined as

$$\mathbf{R}=\begin{bmatrix}0 & 0 & 0 & 0\\ 0 & 0 & 2\kappa & 0\\ 0 & -2\kappa & 0 & 0\\ 0 & 0 & 0 & 0\end{bmatrix} \quad (14)$$

Equation (13) is the finally derived GVRTE for non-participating graded index media. This equation can also be considered as a generalization of the Clausius invariant to Stokes parameters to include polarization. The first component ($I$) of Eq. (13) gives the Clausius invariant. It should be noted that the torsion $\kappa$ is a ray related parameter that may be different for different rays, which indicates the matrix $\mathbf{R}$ is not a physical property of the medium though it appears in the equation like a physical property of the medium. The torsion $\kappa$ along the ray should also be determined during a solution of the transport equation.

It is observed that the torsion of the ray trajectory only affect the $Q$ and $U$ component of the Stokes parameters. If the ray trajectory forms a plane curve, then $\kappa$ is zero, it follows

$$\frac{d}{ds}\left[\frac{\mathbf{I}}{n^2}\right] = 0 \tag{15}$$

If the refractive index distribution is uniform, then Eq. (15) reduces to

$$\frac{d\mathbf{I}}{ds} = 0 \tag{16}$$

This is consistent with the law of conservation of Stokes parameters indicated by the classic VRTE for non-participating media.

*3.2 Participating medium: general conservation*

The transport equation that considering the absorption, emission and scattering process can be simply derived based on the result obtained in Section 3.1. The total variation of Stokes parameters along a differential distance in the ray trajectory in a graded index participating medium can be decomposed into two parts as

$$d\left[\frac{\mathbf{I}}{n^2}\right] = \partial_s\left[\frac{\mathbf{I}}{n^2}\right] + \partial_n\left[\frac{\mathbf{I}}{n^2}\right] \tag{17}$$

where the first partial contribution $\partial_s\left[\mathbf{I}/n^2\right]$ only accounts for the effect of light interaction with medium, namely, the absorption, emission and scattering process, while the second partial contribution $\partial_n\left[\mathbf{I}/n^2\right]$ only accounts for the effect of graded index on the transport process or the geometrical effect.

The first partial contribution $\partial_s\left[\mathbf{I}/n^2\right]$ can be understood as the contribution in condition of a local uniform refractive index, which have been well studied in the VRTE [17]. Hence, this contribution ($\partial_s\left[\mathbf{I}/n^2\right]$) can be written as

$$\partial_s\left[\frac{\mathbf{I}}{n^2}\right] = \left(-\boldsymbol{\kappa}_e \mathbf{I} + \boldsymbol{\kappa}_a \mathbf{I}_b + \frac{1}{4\pi}\int_{4\pi} \mathbf{Z}(\boldsymbol{\Omega}',\boldsymbol{\Omega})\mathbf{I}(s,\boldsymbol{\Omega}')d\Omega\right)\frac{ds}{n^2} \tag{18}$$

in which $\boldsymbol{\kappa}_e$ is the extinction matrix, $\boldsymbol{\kappa}_a$ is the emission matrix, $\mathbf{I}_b$ is the Stokes parameter vector of emission in the refractive medium, and $\mathbf{Z}$ is the scattering phase matrix. These radiative properties are spectrally dependent. The spectral dependence is not explicitly written for brevity. It is noted that the further formulation of the scattering phase matrix $\mathbf{Z}$ based on the scattering matrix $\mathbf{P}$ (such as the Mie scattering matrix) needs some coordinate transformation related to the locally defined Stokes parameters, which will be discussed in detail latter in this section.

As for the second partial contribution $\partial_n[\mathbf{I}/n^2]$, which only account for the geometrical effect. Referring to the result obtained in Section 3.1, $\partial_n[\mathbf{I}/n^2]$ can be written as

$$\partial_n\left[\frac{\mathbf{I}}{n^2}\right] = -\mathbf{R}\,\mathbf{I}\,\frac{\mathrm{d}s}{n^2} \tag{19}$$

Substituting Eq. (18) and Eq. (19) into Eq. (17), it follows

$$n^2\frac{\mathrm{d}}{\mathrm{d}s}\left[\frac{\mathbf{I}}{n^2}\right] + (\boldsymbol{\kappa}_e + \mathbf{R})\,\mathbf{I} = \boldsymbol{\kappa}_a \mathbf{I}_b + \frac{1}{4\pi}\int_{4\pi}\mathbf{Z}(\boldsymbol{\Omega}',\boldsymbol{\Omega})\mathbf{I}(s,\boldsymbol{\Omega}')\mathrm{d}\Omega \tag{20}$$

which is the finally derived GVRTE (the spectral dependence is not explicitly written for brevity). If the refractive index distribution in the medium is uniform, then Eq. (20) reduces to the form of the classic VRTE.

Following the same principle presented above, the scalar radiative transfer equation for light transport in graded index media (GRTE) can also be derived, which reads

$$n^2\frac{\mathrm{d}}{\mathrm{d}s}\left[\frac{I}{n^2}\right] + \kappa_e I = \kappa_a I_b + \frac{\kappa_s}{4\pi}\int_{4\pi}\Phi(\boldsymbol{\Omega}',\boldsymbol{\Omega})I(s,\boldsymbol{\Omega}')\mathrm{d}\Omega \tag{21}$$

where $\kappa_e$, $\kappa_a$, and $\kappa_s$ are the extinction, absorption and scattering coefficient, respectively, $I_b$ is the black body emission intensity in the refractive medium and $\Phi$ is the single scattering phase function.

It is noted that the transport operator derived here [left hand side of Eq. (20)] is consistent with the one derived by Lau and Wartson [18] (proof is given in the Appendix B). However, an explicit formulation of the scattering phase matrix $\mathbf{Z}$ was not given in Ref. [18], which makes the derived equation not readily applicable to analyze radiative transfer in scattering graded index media because the scattering matrix $\mathbf{P}$ is commonly known instead of the scattering phase matrix $\mathbf{Z}$. Furthermore, their derivation used the eikonal approximation following much complicated manipulations and the derived transport equation has not been well verified till its establishment. The result presented here gives an initial mutual verification.

In the following, the explicit formulation of the scattering phase matrix $\mathbf{Z}$ based on the scattering matrix $\mathbf{P}$ is derived. At a specific position along the ray coordinate, the scattering of Stokes parameters from $\boldsymbol{\Omega}'$ to $\boldsymbol{\Omega}$ is schematically shown in Fig. 4. Because rays propagate along different curved ray trajectories for each direction, the local Frenet-Serret frames of the incident direction $\mathbf{\Omega}'$ and the scattered direction $\mathbf{\Omega}$ are different (which are denoted in Fig. 4 as $\mathbf{v}'$-$\mathbf{b}'$-$\mathbf{\Omega}'$ and $\mathbf{v}$-$\mathbf{b}$-$\mathbf{\Omega}$, respectively). To calculate the scattered Stokes parameters, a common reference is needed, which is shown in Fig. 4 as x'-y'-z'. The unit vector of the z' axis ($\mathbf{k}'$) is determined through the binormals of the incident and the scattered direction as $\mathbf{k}' = \dfrac{\mathbf{b} \times \mathbf{b}'}{|\mathbf{b} \times \mathbf{b}'|}$.

In the common reference frame x'-y'-z' (in Fig. 4), the parallel ('||') and perpendicular ('⊥') components of the polarized beam originally defined in the local Frenet-Serret frame are parallel and perpendicular to the meridian plane, respectively. And the polarization components of the incident and the scattered beam are not aligned with the scattering plane $OP'P$, as such, a rotation of Stokes parameters is needed in calculating the scattering phase matrix $\mathbf{Z}$. A transformation of the scattering matrix $\mathbf{P}$ defined based on the scattering plane to the scattering phase matrix $\mathbf{Z}$ can be expressed as

$$\begin{aligned}\mathbf{Z}(\mathbf{\Omega}' \to \mathbf{\Omega}) &= \mathfrak{R}(\phi - \pi)\mathbf{P}(\mathbf{\Omega}' \to \mathbf{\Omega})\mathfrak{R}(\phi') \\ &= \mathfrak{R}(\phi)\mathbf{P}(\mathbf{\Omega}' \to \mathbf{\Omega})\mathfrak{R}(\phi')\end{aligned} \qquad \phi, \phi' \in [0, \pi] \qquad (22)$$

where the rotation matrix $\mathfrak{R}(\phi)$ is defined as

$$\mathfrak{R}(\phi) = \begin{bmatrix} 1 & 0 & 0 & 0 \\ 0 & \cos 2\phi & \sin 2\phi & 0 \\ 0 & -\sin 2\phi & \cos 2\phi & 0 \\ 0 & 0 & 0 & 1 \end{bmatrix} \qquad (23)$$

Based on the triangular relations $\cos 2\phi = 2\cos^2 \phi - 1$ and $\sin 2\phi = 2\sin \phi \cos \phi$, it is easy to show that the knowledge of $\cos \phi$ and $\cos \phi'$ is enough to define the scattering phase matrix. The formulation of $\cos \phi$ and $\cos \phi'$ can be obtained based on vector analysis as

$$\begin{aligned}\cos \phi' &= \mathbf{b}' \bullet \mathbf{n}_{OP'P} \\ \cos \phi &= -\mathbf{b} \bullet \mathbf{n}_{OP'P}\end{aligned} \qquad (24)$$

where $\mathbf{n}_{OP'P} = \dfrac{\mathbf{\Omega}' \times \mathbf{\Omega}}{|\mathbf{\Omega}' \times \mathbf{\Omega}|}$ is the normal of the scattering plane $OP'P$.

It is noted that the transformation given here is very similar to the one given by Chandrasekhar [17] for uniform refractive index media. Even though, there are two major differences, 1) the scattering matrix $\mathbf{P}$ is defined for Stokes vector expressed in right-hand frame in this paper, and 2) the explicit definition of the rotation matrix $\mathfrak{R}$ is related to the local Frenet-Serret frame as shown in Eq. (24).

If it is for the scattering matrix $\mathbf{P}_{LF}$ defined for Stokes vector defined in a left-hand frame, then an additional transformation is needed and the scattering phase matrix can be expressed as

$$\begin{aligned}\mathbf{Z}(\mathbf{r},\mathbf{\Omega}'\rightarrow\mathbf{\Omega}) &= \mathfrak{R}(\phi-\pi)\,\mathbf{T}\,\mathbf{P}_{LF}(\mathbf{r},\mathbf{\Omega}'\rightarrow\mathbf{\Omega})\,\mathbf{T}\,\mathfrak{R}(\phi') \\ &= \mathfrak{R}(\phi)\,\mathbf{T}\,\mathbf{P}_{LF}(\mathbf{r},\mathbf{\Omega}'\rightarrow\mathbf{\Omega})\,\mathbf{T}\,\mathfrak{R}(\phi')\end{aligned} \quad (25)$$

where $\mathbf{T}$ is a transformation matrix to transform Stokes vector defined in the right-hand frame to the left-hand frame, its definition is given in the Appendix A.

*3.3 Boundary conditions*

The GVRTE has been well defined above. The related boundary condition will be discussed in this section. Generally speaking, the boundary condition for the classic VRTE can serve as the boundary condition for the GVRTE. However, a problem still exists because it is commonly difficult to assume the locally defined Stokes vector $\mathbf{I}$ as known. The local Frenet-Serret frame is dependent on the ray trajectory.

From experience, the known Stokes vector at the boundary is often defined based on a fixed experimental frame or based on the surface normal. Hence a transformation is needed to transform the known Stokes vector to the local Frenet-Serret frame. A general transformation can be defined, which is given in the following. A schematic to define the transformation is shown in Fig. 5, where the z- axis denotes the fixed reference. A meridian plane can be defined based on this reference axis and then the parallel and the perpendicular component of the polarized beam can be defined. As is shown, there are two cases [shown in Fig. 5 (a) and (b), respectively] need to be discussed separately.

For the first case [Fig. 5 (a)], $(\mathbf{\Omega}\times\mathbf{v})\cdot\mathbf{e}_z > 0$, a transformation of the Stokes vector defined in the Frenet-Serret frame $\mathbf{I}$ to the fixed reference frame ($\mathbf{I}_G$) is obtained as

$$\mathbf{I}_G = \Re(\phi_G)\mathbf{I}_{FS} \tag{26}$$

and its inverse transform is easily obtained as

$$\mathbf{I} = \Re(-\phi_G)\mathbf{I}_G \tag{27}$$

where $\phi_G$ is a rotation angle between the plane defined by the propagation direction $\mathbf{\Omega}$ and the principal normal $\mathbf{v}$ and the plane defined by $\mathbf{\Omega}$ and the z- axis. Similarly, knowledge of $\cos\phi_G$ is enough to define the rotation, which can be obtained as

$$\cos\phi_G = (\mathbf{\Omega}\times\mathbf{v})\cdot(\mathbf{\Omega}\times\mathbf{e}_z) \tag{28}$$

where $\mathbf{e}_z$ denotes the unit vector of the reference z direction.

For the second case [Fig. 5 (a)], $(\mathbf{\Omega}\times\mathbf{v})\cdot\mathbf{e}_z < 0$, a transformation is obtained as

$$\mathbf{I}_G = \Re(-\phi_G)\mathbf{I} \tag{29}$$

The transformation [Eq. (26) and Eq. (29)] can be cast into one formula as

$$\mathbf{I}_G = \tilde{\Re}(\phi_G)\mathbf{I} \tag{30}$$

where $\tilde{\Re}(\phi_G)$ is a modified rotation matrix given as

$$\tilde{\Re}(\phi_G) = \left\{\frac{1}{2}\big(1+\mathrm{sgn}[(\mathbf{\Omega}\times\mathbf{v})\cdot\mathbf{e}_z]\big)\Re(\phi_G) \right.$$
$$\left. +\big(1-\mathrm{sgn}[(\mathbf{\Omega}\times\mathbf{v})\cdot\mathbf{e}_z]\big)\Re(-\phi_G)\right\} \tag{31}$$

and the unified inverse transform is given as

$$\mathbf{I} = \tilde{\Re}(-\phi_G)\mathbf{I}_G \tag{32}$$

With the transformation defined above, the boundary condition for the Stokes vector defined in the Frenet-Serret frame $\mathbf{I}$ can be transformed to the known Stokes vector $\mathbf{I}_G$ defined in a global fixed reference frame, which can be written as

$$\mathbf{I}(s_0,\mathbf{\Omega}) = \tilde{\Re}(-\phi_G)\mathbf{I}_G(s_0,\mathbf{\Omega}) \tag{33}$$

where $s_0$ is the ray coordinate at the boundary. $\mathbf{I}_G(s_0,\mathbf{\Omega})$ can be considered as the common boundary condition for the VRTE. If the known Stokes vector is given in a left-hand frame denoted as $\mathbf{I}_{G,LF}$, a transformation from left-hand frame to right-hand frame is needed, then Eq. (33) should be written as

$$\mathbf{I}(s_0, \mathbf{\Omega}) = \tilde{\mathfrak{R}}(-\phi_G) \, \mathbf{T} \, \mathbf{I}_{G,LF}(s_0, \mathbf{\Omega}) \tag{34}$$

Equation (33) gives a definition of the boundary condition for the GVRTE based on the boundary condition for the VRTE. A generalized boundary condition for the VRTE which includes Fresnel reflection, Lambertian reflection and collimated beam propagating [22, 23] can be written in the surface normal based reference frame as

$$\begin{aligned}
\mathbf{I}_G(s_0, \mathbf{\Omega}) =& \left[ f_{Fr} \mathbf{\varepsilon}_{Fr}(|\mathbf{n}_w \cdot \mathbf{\Omega}|) + f_{La} \mathbf{\varepsilon}_{La} \right] I_b(\mathbf{r}_w) + \mathbf{I}_{c,G}(s_0) \delta(\mathbf{\Omega} - \mathbf{\Omega}_C) \\
&+ f_{Fr} \mathbf{R}_{Fr}(\mathbf{n}_w \cdot \mathbf{\Omega}) \mathbf{I}_G(s_0, \mathbf{\Omega}'') \\
&+ \frac{f_{La}}{\pi} \left[ \int_{\mathbf{n}_w \cdot \mathbf{\Omega}' > 0} \mathbf{R}_{La} \, \mathbf{I}(s_0, \mathbf{\Omega}') |\mathbf{\Omega}' \cdot \mathbf{n}_w| \, d\Omega' \right]
\end{aligned} \tag{35}$$

where $\mathbf{\varepsilon}_{Fr}$ and $\mathbf{\varepsilon}_{La}$ are the surface emissivity vector of the Fresnel and Lambertian emission, respectively; $f_{Fr}$ and $f_{La} = (1 - f_{Fr})$ is the fraction of Fresnel reflection and diffuse reflection, respectively; $\mathbf{I}_{c,G}$ is the irradiation flux vector of a collimated beam propagating in the direction $\mathbf{\Omega}_C$; $\mathbf{\Omega}''$ is the corresponding incident direction of the current reflected beam of direction $\mathbf{\Omega}$; $\tilde{\psi}$ is the angle between the emission optical plane and the meridian plane of direction $\mathbf{\Omega}$, $\tilde{\psi}''$ is the angle between the optical plane and the meridian plane of the original incident direction of Fresnel reflection $\mathbf{\Omega}''$; $\mathbf{n}_w$ is the unit normal vector of the boundary; $\mathbf{R}_{Fr}$ is the reflection matrix defined by the Fresnel's law. Details on the definition of the boundary condition refer to Ref. [23].

## 4. The transport equation in different forms

The transport equations derived in the previous section, namely, Eq. (13), Eq. (20) and Eq. (21), are formulated in the Lagrangian ray coordinate, which in the Lagrangian form are general and independent of different coordinates system selected. In this section, we consider the different forms of the GVRTE, which include the form for Stokes parameters defined with a fixed reference and the Eulerian forms in ray coordinate and in several common orthogonal coordinate systems.

*4.1 Forms for Stokes parameters defined with a fixed reference*

The Stokes vector that appears in the GVRTE derived in Section 3 is defined in the local Frenet-Serret frame. Generally, the local Frenet-Serret frame varies with different position and different propagating direction in the graded index medium, which makes the solved Stokes vector difficult to be interpreted due to the variable reference frame. A question is thus put forward, is it possible to formulate the GVRTE in a global fixed reference frame for Stokes vector? The question is yes and the form of the GVRTE in a global fixed reference frame for Stokes vector is derived and discussed in this section.

In Section 3.3, a transformation is obtained between the Stokes vector defined in the local Frenet-Serret frame and the Stokes vector defined in a global fixed frame. Substitute the inverse transformation, namely, Eq. (32) into the GVRTE [Eq. (20)] yields

$$\tilde{\mathfrak{R}}(-\phi_G)n^2 \frac{\mathrm{d}}{\mathrm{d}s}\left[\frac{\mathbf{I}_G}{n^2}\right] + \frac{\mathrm{d}}{\mathrm{d}s}\left[\tilde{\mathfrak{R}}(-\phi_G)\right]\mathbf{I}_G$$
$$+ \left(\boldsymbol{\kappa}_e + \mathbf{R}\right)\tilde{\mathfrak{R}}(-\phi_G)\mathbf{I}_G \qquad (36)$$
$$= \boldsymbol{\kappa}_a \mathbf{I}_{b,FS} + \frac{1}{4\pi}\int_{4\pi} \mathbf{Z}(\boldsymbol{\Omega}',\boldsymbol{\Omega})\tilde{\mathfrak{R}}(-\phi_G')\mathbf{I}_G(s,\boldsymbol{\Omega}')\mathrm{d}\Omega$$

Multiplying Eq. (36) with $\tilde{\mathfrak{R}}(\phi)$ which is the inverse of $\tilde{\mathfrak{R}}(-\phi_G)$, it follows

$$n^2 \frac{\mathrm{d}}{\mathrm{d}s}\left[\frac{\mathbf{I}_G}{n^2}\right] + \tilde{\mathfrak{R}}(\phi_G)\frac{\mathrm{d}}{\mathrm{d}s}\left[\tilde{\mathfrak{R}}(-\phi_G)\right]\mathbf{I}_G$$
$$+ \tilde{\mathfrak{R}}(\phi_G)\left(\boldsymbol{\kappa}_e + \mathbf{R}\right)\tilde{\mathfrak{R}}(-\phi_G)\mathbf{I}_G$$
$$= \tilde{\mathfrak{R}}(\phi_G)\boldsymbol{\kappa}_a\tilde{\mathfrak{R}}(-\phi_G)\mathbf{I}_{b,G} \qquad (37)$$
$$+ \frac{1}{4\pi}\int_{4\pi}\tilde{\mathfrak{R}}(\phi_G)\mathbf{Z}(\boldsymbol{\Omega}',\boldsymbol{\Omega})\tilde{\mathfrak{R}}(-\phi_G')\mathbf{I}_G(s,\boldsymbol{\Omega}')\mathrm{d}\Omega$$

which can be further written as

$$n^2 \frac{\mathrm{d}}{\mathrm{d}s}\left[\frac{\mathbf{I}_G}{n^2}\right] + \left[\tilde{\mathfrak{R}}(\phi_G)\frac{\mathrm{d}\tilde{\mathfrak{R}}(-\phi_G)}{\mathrm{d}s} + \left(\underset{\sim}{\boldsymbol{\kappa}}_e + \underset{\sim}{\mathbf{R}}\right)\right]\mathbf{I}_G$$
$$= \underset{\sim}{\boldsymbol{\kappa}}_a\mathbf{I}_{b,G} + \frac{1}{4\pi}\int_{4\pi}\underset{\sim}{\mathbf{Z}}(\boldsymbol{\Omega}',\boldsymbol{\Omega})\mathbf{I}_G(s,\boldsymbol{\Omega}')\mathrm{d}\Omega \qquad (38)$$

where $\underset{\sim}{\boldsymbol{\kappa}}_e$, $\underset{\sim}{\mathbf{R}}$, $\underset{\sim}{\boldsymbol{\kappa}}_a$ and $\underset{\sim}{\mathbf{Z}}$ are the corresponding modified polarized radiative properties defined as

$$\underset{\sim}{\boldsymbol{\kappa}}_e = \tilde{\Re}(\phi_G)\,\boldsymbol{\kappa}_e\,\tilde{\Re}(-\phi_G) \tag{39a}$$

$$\underset{\sim}{\mathbf{R}} = \tilde{\Re}(\phi_G)\,\mathbf{R}\,\tilde{\Re}(-\phi_G) \tag{39b}$$

$$\underset{\sim}{\boldsymbol{\kappa}}_a = \tilde{\Re}(\phi_G)\boldsymbol{\kappa}_a\tilde{\Re}(-\phi_G) \tag{39c}$$

$$\underset{\sim}{\mathbf{Z}}(\boldsymbol{\Omega}',\boldsymbol{\Omega}) = \tilde{\Re}(\phi_G)\mathbf{Z}(\boldsymbol{\Omega}',\boldsymbol{\Omega})\tilde{\Re}(-\phi_G') \tag{39d}$$

Equation (38) is the finally derived equation of the GVRTE in a global fixed reference frame for Stokes vector. It is seen that an additional term [second term in Eq. (38)] appears besides the modified radiative properties, which makes this equation more complicated than the original GVRTE. If the medium is with uniform refractive index distribution, then the rotation matrix equals to identity matrix, and Eq. (38) will reduce to the classic VRTE.

*4.2 Eulerian forms in ray coordinate and in common orthogonal coordinate systems*

**4.2.1 Eulerian form in ray coordinate**

To adapt the transport equations to transient process, the Lagrangian streaming operator, $d/ds$, is further expanded in spatial and temporal space and give a Eulerian form in ray coordinate as

$$\frac{d}{ds} = \frac{dt}{ds}\frac{\partial}{\partial t} + \frac{\partial}{\partial s} = \frac{n}{c}\frac{\partial}{\partial t} + \frac{\partial}{\partial s} \tag{40}$$

where $c$ is the speed of light in vacuum. Substituting the expanded streaming operator Eq. (40) to Eq. (20), the Eulerian form of the GVRTE in ray coordinate is obtained as

$$\frac{n}{c}\frac{\partial \mathbf{I}}{\partial t} + n^2\frac{\partial}{\partial s}\left[\frac{\mathbf{I}}{n^2}\right] + \left(\boldsymbol{\kappa}_e + \mathbf{R}\right)\mathbf{I} = \boldsymbol{\kappa}_a\mathbf{I}_b + \frac{1}{4\pi}\int_{4\pi}\mathbf{Z}(\boldsymbol{\Omega}',\boldsymbol{\Omega})\mathbf{I}(t,s,\boldsymbol{\Omega}')d\Omega \tag{41}$$

To solve the Eulerian form of the GVRTE in ray coordinate, a ray tracing technique is essential, which is difficult for graded index media and also usually inefficient for scattering problem. According to previous studies on the solution of scalar radiative transfer in graded index media, numerical methods based on discretizing the Eulerian forms of the GRTE in some orthogonal coordinate systems is very efficient [24-28], in which ray tracing is avoided. Hence to further formulate the Eulerian form of the GVRTE in common orthogonal coordinate systems is very appealing for developing efficient solution techniques for polarized radiative transfer in graded index media.

In the following, the Eulerian form of the GVRTE [Eq. (41)] is formulated to several common orthogonal coordinate systems.

**4.2.2 Eulerian forms in common orthogonal coordinate systems**

Here in this section, a compilation of different forms of the GVRTE for several common orthogonal coordinate systems are presented. As can be seen from Eq. (41), there are two major tasks to do in order to formulate the GVRTE to other coordinate systems, one is to expand the Eulerian streaming operator $\partial/\partial s$ in that coordinate system, the other is to calculate the torsion $\kappa$ of ray trajectory which appear in the rotation matrix $\mathbf{R}$ in the related coordinate system. Using the Eulerian formulation of the GVRTE, the Stokes parameters is a function of time, spatial coordinates and angular coordinates, namely, $\mathbf{I}(t,\mathbf{r},\mathbf{\Omega})$, where $\mathbf{r}$ is the spatial vector. In the following description, the arguments of $\mathbf{I}(t,\mathbf{r},\mathbf{\Omega})$ is omitted for brevity in case there is no ambiguity.

The expanding of the Eulerian streaming operator $\partial/\partial s$ should consider the geometrical effect caused by the graded index distribution. The curved ray trajectory is determined based on the ray equation. The expanding of $\partial/\partial s$ has been well studied for the GRTE [26, 29, 30], which can also be applied to the GVRTE. Hence, the detailed manipulation on the derivations will not be repeated here and the major results are taken directly.

For the Cartesian coordinates system $x$-$y$-$z$-$\theta$-$\varphi$ (defined in Fig. 6), the Eulerian streaming operator $\partial/\partial s$ can be expanded as [26]

$$\frac{\partial}{\partial s} = \mathbf{\Omega}\cdot\nabla + \frac{1}{\sin\theta}\left\{\left[(\xi\mathbf{\Omega}-\mathbf{k})\cdot\frac{\nabla n}{n}\right]\frac{\partial}{\partial\theta} + \left[\mathbf{s}_1\cdot\frac{\nabla n}{n}\right]\frac{\partial}{\partial\varphi}\right\} \tag{42}$$

where $\mathbf{\Omega} = \mathbf{i}\mu + \mathbf{j}\eta + \mathbf{k}\xi$ is the propagation direction vector of the ray, $\mu = \sin\theta\cos\varphi$, $\eta = \sin\theta\sin\varphi$ and $\xi = \cos\theta$, which are the direction cosines of the beam direction, $\nabla = \mathbf{i}\partial/\partial x + \mathbf{j}\partial/\partial y + \mathbf{k}\partial/\partial z$ is the gradient operator, $\mathbf{s}_1 = -\mathbf{i}\sin\varphi + \mathbf{j}\cos\varphi$ is an auxiliary vector. Furthermore, it is easy to verify that

$$n^2\frac{\partial}{\partial s}\left[\frac{\mathbf{I}}{n^2}\right] = \frac{\partial\mathbf{I}}{\partial s} - \frac{2}{n}\frac{\partial n}{\partial s}\mathbf{I} \tag{43}$$

Hence the term $n^2 \partial (\mathbf{I}/n^2)/\partial s$ in Eq. (41) can be obtained in non-conservative form as

$$n^2 \frac{\partial}{\partial s}\left[\frac{\mathbf{I}}{n^2}\right] = \mathbf{\Omega} \cdot \nabla \mathbf{I}$$
$$+ \frac{1}{\sin\theta}\left\{\left[(\xi\mathbf{\Omega} - \mathbf{k}) \cdot \frac{\nabla n}{n}\right]\frac{\partial \mathbf{I}}{\partial \theta} + \left(\mathbf{s}_1 \cdot \frac{\nabla n}{n}\right)\frac{\partial \mathbf{I}}{\partial \varphi}\right\} \quad (44)$$
$$- 2\left(\mathbf{\Omega} \cdot \frac{\nabla n}{n}\right)\mathbf{I}$$

and in conservative form as

$$n^2 \frac{\partial}{\partial s}\left[\frac{\mathbf{I}}{n^2}\right] = \mathbf{\Omega} \cdot \nabla \mathbf{I} + \frac{1}{\sin\theta}\left\{\frac{\partial}{\partial \theta}\left[\left\{(\xi\mathbf{\Omega} - \mathbf{k}) \cdot \frac{\nabla n}{n}\right\}\mathbf{I}\right] + \frac{\partial}{\partial \varphi}\left[\left(\mathbf{s}_1 \cdot \frac{\nabla n}{n}\right)\mathbf{I}\right]\right\} \quad (45)$$

The torsion of the ray trajectory, which can be obtained based on the ray equation and the Frenet-Serret formula and can be written as [9]

$$\kappa = -\mathbf{b} \cdot \frac{\partial \mathbf{v}}{\partial s} = -\mathbf{b} \cdot \frac{\partial}{\partial s}\left[\frac{(\boldsymbol{\delta} - \mathbf{\Omega}\mathbf{\Omega})}{K} \cdot \frac{\nabla n}{n}\right] \quad (46)$$

where $\boldsymbol{\delta}$ is the unit tensor, the binormal $\mathbf{b} = (\mathbf{\Omega} \times \nabla n/n)/K$ and $K = |\mathbf{\Omega} \times \nabla n/n|$ is the local curvature of the ray. Equation (46) can be further written as

$$\kappa = \frac{\mathbf{\Omega} \times \nabla n}{|\mathbf{\Omega} \times \nabla n|} \cdot \frac{\partial}{\partial s}\left[\frac{(\mathbf{\Omega}\mathbf{\Omega} - \boldsymbol{\delta}) \cdot \nabla n}{|\mathbf{\Omega} \times \nabla n|}\right] \quad (47)$$

It is seen that the explicit formula of the torsion in Cartesian coordinates still require the expansion of the Eulerian streaming operator $\partial/\partial s$, which is given by Eq. (42). Hence the final form of the formula for the torsion is rather complicated. Substituting Eq. (45) and Eq. (46) into Eq. (41), the Cartesian form of the GVRTE is then obtained (in conservative form) as

$$\frac{n}{c}\frac{\partial \mathbf{I}}{\partial t} + \mathbf{\Omega} \cdot \nabla \mathbf{I}$$
$$+ \frac{1}{\sin\theta}\left\{\frac{\partial}{\partial \theta}\left[\left\{(\xi\mathbf{\Omega} - \mathbf{k}) \cdot \frac{\nabla n}{n}\right\}\mathbf{I}\right] + \frac{\partial}{\partial \varphi}\left[\left(\mathbf{s}_1 \cdot \frac{\nabla n}{n}\right)\mathbf{I}\right]\right\} \quad (48)$$
$$+ (\kappa_e + R)\mathbf{I} = \kappa_a \mathbf{I}_b + \frac{1}{4\pi}\int_{4\pi} \mathbf{Z}(\mathbf{\Omega}', \mathbf{\Omega})\mathbf{I}(t, \mathbf{r}, \mathbf{\Omega}')d\Omega$$

As shown above, with the knowledge of the formula for the Eulerian streaming operator $\partial/\partial s$ and the second term of the Eulerian form GVRTE in different coordinate systems, it is easy to obtain the variant form of the GVRTE for different coordinate systems. Hence only the explicit formulas of these relations are presented in the following description.

Two kinds of cylindrical coordinates systems ($\rho$-$\Psi$-$z$-$\theta$-$\varphi$) are considered, one is the traditional cylindrical coordinates system [31] [defined in Fig. 7(a)] and the other is a new cylindrical coordinates system [defined in Fig. 7(b)] introduced recently [30, 32] which is advantageous in dealing with specular reflection/refraction at the cylindrical boundary. For the traditional cylindrical coordinates system [Fig. 7(a)], the Eulerian streaming operator $\partial/\partial s$ can be expanded as [29]

$$\frac{\partial}{\partial s} = \mathbf{\Omega} \cdot \nabla - \frac{\eta}{\rho}\frac{\partial}{\partial \varphi} + \frac{1}{\sin\theta}\left\{\left[(\xi\mathbf{\Omega} - \mathbf{e}_z) \cdot \frac{\nabla n}{n}\right]\frac{\partial}{\partial \theta} + \left[\mathbf{s}_1 \cdot \frac{\nabla n}{n}\right]\frac{\partial}{\partial \varphi}\right\} \tag{49}$$

where $\mathbf{\Omega} = \mu\mathbf{e}_\rho + \eta\mathbf{e}_\Psi + \xi\mathbf{e}_z$ is the local direction vector of the beam, $\mathbf{e}_\rho$, $\mathbf{e}_\Psi$ and $\mathbf{e}_z$ are the unit coordinate vector, $\nabla = \mathbf{e}_\rho \partial/\partial\rho + \mathbf{e}_\Psi \rho^{-1}\partial/\partial\Psi + \mathbf{e}_z \partial/\partial z$ is the gradient operator, $\mathbf{s}_1 = -\mathbf{e}_\rho \sin\varphi + \mathbf{e}_\Psi \cos\varphi$ is an auxiliary vector. Following the similar procedure, the second term in Eq. (41) can be obtained in the non-conservative form as

$$n^2\frac{\partial}{\partial s}\left[\frac{\mathbf{I}}{n^2}\right] = \mathbf{\Omega} \cdot \nabla \mathbf{I} - \frac{\eta}{\rho}\frac{\partial \mathbf{I}}{\partial \varphi}$$
$$+ \frac{1}{\sin\theta}\left\{\left[(\xi\mathbf{\Omega} - \mathbf{e}_z) \cdot \frac{\nabla n}{n}\right]\frac{\partial \mathbf{I}}{\partial \theta} + \left[\mathbf{s}_1 \cdot \frac{\nabla n}{n}\right]\frac{\partial \mathbf{I}}{\partial \varphi}\right\} \tag{50}$$
$$- 2\left(\mathbf{\Omega} \cdot \frac{\nabla n}{n}\right)\mathbf{I}$$

and in the conservative form as

$$n^2\frac{\partial}{\partial s}\left[\frac{\mathbf{I}}{n^2}\right] = \mathbf{\Omega} \cdot \tilde{\nabla}\mathbf{I} - \frac{1}{\rho}\frac{\partial \eta \mathbf{I}}{\partial \varphi}$$
$$+ \frac{1}{\sin\theta}\left\{\frac{\partial}{\partial\theta}\left[\left((\xi\mathbf{\Omega} - \mathbf{e}_z) \cdot \frac{\nabla n}{n}\right)\mathbf{I}\right] + \frac{\partial}{\partial\varphi}\left[\left(\mathbf{s}_1 \cdot \frac{\nabla n}{n}\right)\mathbf{I}\right]\right\} \tag{51}$$

where $\tilde{\nabla}(\bullet) = \mathbf{e}_\rho \rho^{-1}\partial(\rho\bullet)/\partial\rho + \mathbf{e}_\Psi \rho^{-1}\partial(\bullet)/\partial\Psi + \mathbf{e}_z \rho^{-1}\partial(\bullet)/\partial z$ is a modified gradient operator. The formula for the torsion can be obtained by substitution of the expanded streaming operator [Eq. (49)] into Eq. (47).

For the new cylindrical coordinates system [Fig. 7(b)], the Eulerian streaming operator $\partial/\partial s$ can be expanded as [30]

$$\frac{\partial}{\partial s} = \mathbf{\Omega} \cdot \nabla - \mu \frac{\cos\varphi}{\rho} \frac{\partial}{\partial \theta} + \xi \frac{\sin\varphi \cos\varphi}{\rho} \frac{\partial}{\partial \varphi} \\ + \frac{1}{\sin\theta} \left\{ \left[ (\xi\mathbf{\Omega} - \mathbf{e}_\rho) \cdot \frac{\nabla n}{n} \right] \frac{\partial}{\partial \theta} + \left[ \mathbf{s}_1 \cdot \frac{\nabla n}{n} \right] \frac{\partial}{\partial \varphi} \right\}$$
(52)

where $\mathbf{\Omega} = \mu \mathbf{e}_\Psi + \eta \mathbf{e}_z + \xi \mathbf{e}_\rho$ is the local direction vector of the beam, and the gradient operator is given as $\nabla = \mathbf{e}_\Psi \rho^{-1} \partial/\partial \Psi + \mathbf{e}_z \partial/\partial z + \mathbf{e}_\rho \partial/\partial \rho$, $\mathbf{s}_1 = -\sin\varphi \mathbf{e}_\Psi + \cos\varphi \mathbf{e}_z$ is an auxiliary vector. The second term in Eq. (41) can be obtained in the non-conservative form as

$$n^2 \frac{\partial}{\partial s}\left[\frac{\mathbf{I}}{n^2}\right] = \mathbf{\Omega} \cdot \nabla \mathbf{I} - \mu \frac{\cos\varphi}{\rho} \frac{\partial \mathbf{I}}{\partial \theta} + \xi \frac{\sin\varphi \cos\varphi}{\rho} \frac{\partial \mathbf{I}}{\partial \varphi} \\ + \frac{1}{\sin\theta} \left\{ \left[ (\xi\mathbf{\Omega} - \mathbf{e}_\rho) \cdot \frac{\nabla n}{n} \right] \frac{\partial \mathbf{I}}{\partial \theta} + \left[ \mathbf{s}_1 \cdot \frac{\nabla n}{n} \right] \frac{\partial \mathbf{I}}{\partial \varphi} \right\} \\ - 2\left( \mathbf{\Omega} \cdot \frac{\nabla n}{n} \right) \mathbf{I}$$
(53)

and in the conservative form as

$$n^2 \frac{\partial}{\partial s}\left[\frac{\mathbf{I}}{n^2}\right] = \mathbf{\Omega} \cdot \tilde{\nabla} \mathbf{I} - \frac{1}{\rho \sin\theta} \frac{\partial}{\partial \theta}[\mu^2 \mathbf{I}] + \frac{1}{\rho} \frac{\partial}{\partial \varphi}[\xi \sin\varphi \cos\varphi \mathbf{I}] \\ + \frac{1}{\sin\theta} \left\{ \frac{\partial}{\partial \theta}\left[ \left( (\xi\mathbf{\Omega} - \mathbf{e}_z) \cdot \frac{\nabla n}{n} \right) \mathbf{I} \right] + \frac{\partial}{\partial \varphi}\left[ \left( \mathbf{s}_1 \cdot \frac{\nabla n}{n} \right) \mathbf{I} \right] \right\}$$
(54)

where $\tilde{\nabla}(\bullet) = \mathbf{e}_\Psi \rho^{-1} \partial(\bullet)/\partial \Psi + \mathbf{e}_z \partial(\bullet)/\partial z + \mathbf{e}_\rho \rho^{-1} \partial(\rho \bullet)/\partial \rho$ is a modified gradient operator.

For the spherical coordinate system ($\Theta$-$\Psi$-$\rho$-$\theta$-$\varphi$) (defined in Fig. 8), the Eulerian streaming operator $\partial/\partial s$ can be expanded as [29]

$$\frac{\partial}{\partial s} = \mathbf{\Omega} \cdot \nabla - \frac{\sin\theta}{\rho} \frac{\partial}{\partial \theta} - \frac{\eta \cot\Theta}{\rho} \frac{\partial}{\partial \varphi} \\ + \frac{1}{\sin\theta} \left\{ \left[ (\xi\mathbf{\Omega} - \mathbf{e}_\rho) \cdot \frac{\nabla n}{n} \right] \frac{\partial}{\partial \theta} + \left[ \mathbf{s}_1 \cdot \frac{\nabla n}{n} \right] \frac{\partial}{\partial \varphi} \right\}$$
(55)

where $\mathbf{\Omega} = \mu \mathbf{e}_\Theta + \eta \mathbf{e}_\Psi + \xi \mathbf{e}_\rho$ is the local direction vector of the beam, the gradient operator is defined as $\nabla = \mathbf{e}_\Theta \rho^{-1} \partial/\partial \Theta + \mathbf{e}_\Psi (\rho \sin\Theta)^{-1} \partial/\partial \Psi + \mathbf{e}_\rho \partial/\partial \rho$, $\mathbf{s}_1 = -\sin\varphi \mathbf{e}_\Theta + \cos\varphi \mathbf{e}_\Psi$ is an auxiliary vector. The second term in Eq. (41) can be obtained in the non-conservative form as

$$n^2 \frac{\partial}{\partial s}\left[\frac{\mathbf{I}}{n^2}\right] = \mathbf{\Omega} \cdot \nabla \mathbf{I} - \frac{\sin\theta}{\rho}\frac{\partial \mathbf{I}}{\partial \theta} - \frac{\eta \cot\Theta}{\rho}\frac{\partial \mathbf{I}}{\partial \varphi}$$
$$+ \frac{1}{\sin\theta}\left\{\left[(\xi\mathbf{\Omega} - \mathbf{e}_\rho) \cdot \frac{\nabla n}{n}\right]\frac{\partial \mathbf{I}}{\partial \theta} + \left[\mathbf{s}_1 \cdot \frac{\nabla n}{n}\right]\frac{\partial \mathbf{I}}{\partial \varphi}\right\} \quad (56)$$
$$- 2\left(\mathbf{\Omega} \cdot \frac{\nabla n}{n}\right)\mathbf{I}$$

and in the conservative form as

$$n^2 \frac{\partial}{\partial s}\left[\frac{\mathbf{I}}{n^2}\right] = \mathbf{\Omega} \cdot \tilde{\nabla}\mathbf{I} - \frac{1}{\rho\sin\theta}\frac{\partial \sin^2\theta \mathbf{I}}{\partial \theta} - \frac{\cot\Theta}{\rho}\frac{\partial \eta\mathbf{I}}{\partial \varphi}$$
$$+ \frac{1}{\sin\theta}\left\{\frac{\partial}{\partial\theta}\left[\left[(\xi\mathbf{\Omega} - \mathbf{e}_\rho) \cdot \frac{\nabla n}{n}\right]\mathbf{I}\right] + \frac{\partial}{\partial\varphi}\left[\left(\mathbf{s}_1 \cdot \frac{\nabla n}{n}\right)\mathbf{I}\right]\right\} \quad (57)$$

where $\tilde{\nabla}(\bullet) = \mathbf{e}_\Theta (\rho\sin\Theta)^{-1} \partial(\sin\Theta \bullet)/\partial\Theta + \mathbf{e}_\Psi (\rho\sin\Theta)^{-1} \partial/\partial\Psi + \mathbf{e}_\rho \rho^{-2} \partial(\rho^2 \bullet)/\partial\rho$ is a modified gradient operator.

It is noted that the GVRTE in different coordinate systems presented above can serve as the basis for developing numerical techniques for solving the polarized radiative transfer problems in graded index media.

## 5. Conclusions

The vector radiative transfer equation for polarized light transport in absorption, emission and scattering graded index media is briefly derived. The derivation is based on the analysis of the conservation quantities for polarized light transport along curved trajectory and a novel approach. The derived GVRTE can be considered as a generalization of the VRTE for uniform refractive index media. Several variant forms of the transport equations are also presented, which include the forms for Stokes parameters defined with a fixed reference and the Eulerian forms in the ray coordinate and in several common orthogonal coordinate systems (Cartesian, cylindrical and spherical coordinate systems).

The GVRTE provides a sound basis for polarized radiative transfer in graded index media. It and its variant forms can be taken as basis for study the process of polarized radiative transfer in graded index media, such as in the fields of remote sensing, atmospheric optics and biomedical optical tomography. In the related applications where the medium is with inhomogeneous refractive index distribution, the GVRTE based approach is considered to give better predictions and the predictions based on the VRTE may fail.

To solve the polarized radiative transfer in graded index media, there are basically two groups of numerical methods, 1) methods based on ray tracing and 2) methods based on partial differential equation (PDE) discretization in which ray tracing is avoided. The Eulerian form of the GVRTE in the ray coordinate can serve as a basis for the solution methods using ray tracing techniques. The Eulerian form of the GVRTE in common orthogonal coordinate systems can serve as the basis for developing PDE based solution methods for polarized radiative transfer in graded index media.

Generally, the GVRTE and its variant forms show rather complicated forms. The torsion of the ray trajectory needs to be obtained before a solution, which requires much additional efforts. It will not be a trivial work to extend the previously developed solution technique designed for the GRTE or the VRTE to the solution of the GVRTE. Much effort on the solution technique is required to the application of the GVRTE to real problems.

**Acknowledgements**

The supports of this work by the National Natural Science Foundation of China (No.: 51076038, 50906017), the Development Program for Outstanding Young Teachers in Harbin Institute of Technology (HITQNJS.2009.020), and the Program for Changjiang Scholars and Innovative Research Team in University (IRT0914) are gratefully acknowledged.

**Appendix**

*A. Stokes parameters defined in left-hand frame and right-hand frame*

The Stokes parameters can be defined in the left-hand and the right-hand frame. The relation between the two types of definitions is presented here. The definition of variables for the left-hand frame is shown in Fig. 9. The definition based on the long axis and that based on the short axis is shown in Fig. 9(a) and 9(b), respectively. A comparison of variables defined in the two types definition, namely, the right-hand frame (Fig. 1) and the left-hand frame Fig. 9(a), reveals the following relations

$$\psi' = \frac{\pi}{2} - \psi, \quad \tan\chi' = \frac{1}{\tan\chi} \tag{58}$$

where $\psi'$ and $\chi'$ denote the orientation angle and the ellipticity angle of the polarization ellipse, respectively, in the left-hand frame. Base on the triangle relations as follows

$$\cos^2\alpha = \frac{1}{1+\tan^2\alpha}, \quad \cos 2\alpha = \frac{1-\tan^2\alpha}{1+\tan^2\alpha}, \quad \sin 2\alpha = \frac{2\tan\alpha}{1+\tan^2\alpha} \tag{59}$$

It yields

$$\cos 2\chi' = \frac{1-\tan^2\chi'}{1+\tan^2\chi'} = -\frac{1-\tan^2\chi}{1+\tan^2\chi} = -\cos 2\chi \tag{60}$$

$$\sin 2\chi' = \frac{2\tan\chi'}{1+\tan^2\chi'} = \frac{2\tan\chi}{1+\tan^2\chi} = \sin 2\chi \tag{61}$$

Furthermore

$$\cos\left[2\left(\frac{\pi}{2}-\psi\right)\right] = -\cos(2\psi), \quad \sin\left[2\left(\frac{\pi}{2}-\psi\right)\right] = \sin(2\psi) \tag{62}$$

Hence the transformation from the right-hand frame Stokes vector to the left-hand Stokes vector can be defined as

$$\mathbf{I'} = \begin{pmatrix} I \\ I\cos 2\chi'\cos 2\psi' \\ I\cos 2\chi'\sin 2\psi' \\ I\sin 2\chi' \end{pmatrix} = \begin{pmatrix} I \\ I\cos 2\chi\cos 2\psi \\ -I\cos 2\chi\sin 2\psi \\ I\sin 2\chi \end{pmatrix} = \mathbf{T}\mathbf{I} \tag{63}$$

where the transform matrix $\mathbf{T}$ is defined as

$$\mathbf{T} = \begin{bmatrix} 1 & & & \\ & 1 & & \\ & & -1 & \\ & & & 1 \end{bmatrix} \tag{64}$$

It is easy to show that the inverse transform is $\mathbf{I} = \mathbf{T}\mathbf{I'}$.

As for the definition based on the short axis shown in Fig. 9 (b), the orientation angle and the ellipticity angle is related to that in the right-hand frame as

$$\psi' = \pi - \psi, \quad \tan\chi' = \tan\chi \tag{65}$$

Hence

$$\cos 2\chi' = \cos 2\chi \tag{66}$$

$$\sin 2\chi' = \sin 2\chi \tag{67}$$

Furthermore,

$$\cos[2(\pi-\psi)] = \cos(2\psi), \quad \sin[2(\pi-\psi)] = -\sin(2\psi) \tag{68}$$

It is seen that the same transformation as given by Eq. (63) is obtained.

Similarly, it is also easy to prove that the Stokes parameters is invariant under the transformation, $\psi' = \psi + \pi/2$ and $\tan\chi' = 1/\tan\chi$. This indicates the Stokes parameters have the same value for the definitions under the same reference frame (left or right), whether or not the long axis based definition or the short axis based definition is selected.

It is noted that the value of the Stokes vector defined in the left and the right frame are different. Hence transformation [Eq. (63)] is needed if the Stokes vector used in the formula are from different definition. With the knowledge of the transformation of Stokes parameters, the scattering matrix defined in the left-hand frame $\mathbf{P}_{LF}$ can be transformed to the right-hand frame $\mathbf{P}$ as

$$\mathbf{P}(\mathbf{\Omega}' \to \mathbf{\Omega}) = \mathbf{T}\,\mathbf{P}_{LF}(\mathbf{\Omega}' \to \mathbf{\Omega})\,\mathbf{T} \tag{69}$$

and the inverse transform is

$$\mathbf{P}_{LF}(\mathbf{\Omega}' \to \mathbf{\Omega}) = \mathbf{T}\,\mathbf{P}(\mathbf{\Omega}' \to \mathbf{\Omega})\,\mathbf{T} \tag{70}$$

### B. Proof on the consistence of the transport operator derived here with Ref. [18]

The transport equation obtained by Lau and Warton [18] is given as

$$\frac{d}{ds}\mathbf{I} + \left(\frac{1}{L} + \mathbf{R}\right)\mathbf{I} = \mathbf{B} \tag{71}$$

where $\mathbf{B}$ is the source term account for in scattering and $1/L$ is a scalar given as

$$\frac{1}{L} = \frac{1}{l} - \frac{d}{ds}\ln n^2 \tag{72}$$

in which $l$ is the mean free path of photon and $l^{-1}$ should be the same as the extinction coefficient $\kappa_e$. By noticing the following relation

$$-\frac{d}{ds}\left[\ln n^2\right] = \frac{d}{ds}\left[\ln \frac{1}{n^2}\right] = n^2 \frac{d}{ds}\left[\frac{1}{n^2}\right] \tag{73}$$

the transport operator [left hand side of Eq. (71)] can be rewritten as

$$\begin{aligned}
&\frac{d}{ds}\mathbf{I} + \left(\kappa_e - \frac{d}{ds}\ln n_r^2 + \mathbf{R}\right)\mathbf{I} \\
&= \frac{d}{ds}\mathbf{I} + n^2 \frac{d}{ds}\left[\frac{1}{n^2}\right]\mathbf{I} + \kappa_e \mathbf{I} + \mathbf{R}\mathbf{I} \\
&= n^2 \frac{d}{ds}\left[\frac{\mathbf{I}}{n^2}\right] + (\kappa_e + \mathbf{R})\mathbf{I}
\end{aligned} \tag{74}$$

As compared to the transport operator derived in this paper [Eq. (20)], the difference lies in the extinction term. A scalar extinction coefficient is used in the work of Lau and Warton, which can be considered a very simplified case of the extinction matrix $\mathbf{\kappa}_e$ used in the GVRTE. In this understanding, the transport operator derived in this paper is consistent with that derived in Ref. [18].

**Figure Captions**

**Fig. 1.** Definition of variables for the polarization ellipse.

**Fig. 2**. Definition of polarization ellipse in the local Frenet-Serret frame along the curved ray trajectory.

**Fig. 3.** Schematic of the two layer media to derive the relation between the energy flux and the transport intensity.

**Fig. 4.** Geometry for the definition of polarization reference frame and the definition of angles in the transformation of scattering phase matrix.

**Fig. 5.** Schematic to define a transformation of Stokes vector from the local Frenet-Serret frame to a fixed frame. **(a)** Case 1, $(\mathbf{\Omega} \times \mathbf{v}) \cdot \mathbf{e}_z > 0$, **(b)** Case 2, $(\mathbf{\Omega} \times \mathbf{v}) \cdot \mathbf{e}_z < 0$.

**Fig. 6.** Definition of the Cartesian coordinate system.

**Fig. 7.** Definition of the cylindrical coordinate system. **(a)** traditional cylindrical coordinate system, **(b)** new cylindrical coordinate system.

**Fig. 8.** Definition of the spherical coordinate system.

**Fig. 9.** Definition of variables for the polarization ellipse in the left-hand frame based on different axis. **(a)** Case 1, based on the short axis, **(b)** Case 2, based on the long axis.

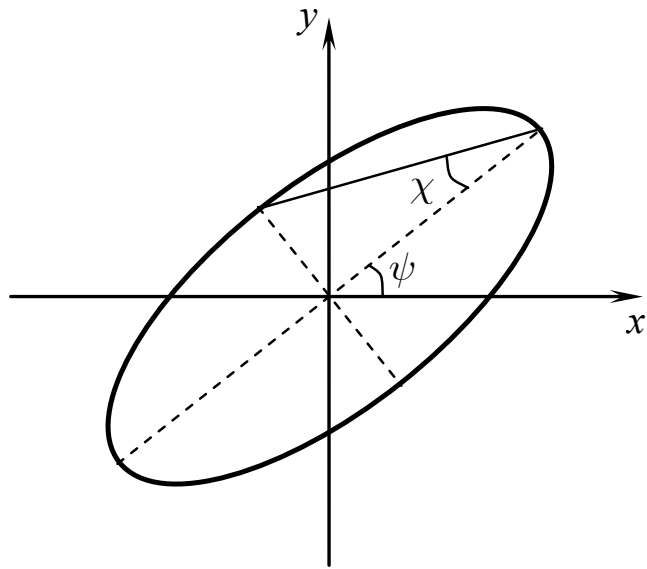

**Fig. 1.** Definitions of variables for the polarization ellipse.

**Authors: Zhao, Tan and Liu**

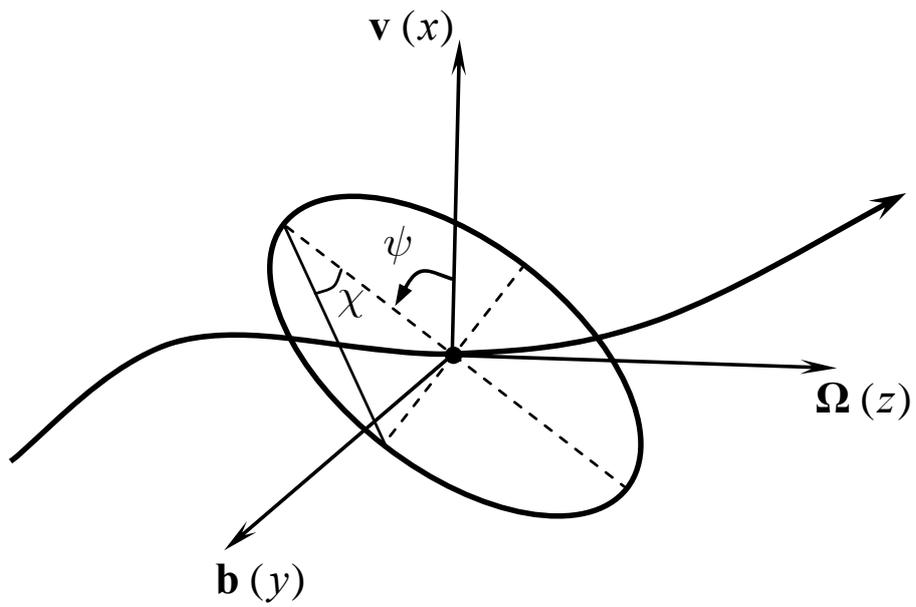

**Fig. 2.** Definitions of variables in the local Frenet-Serret frame along the curved ray trajectory.

**Authors: Zhao, Tan and Liu**

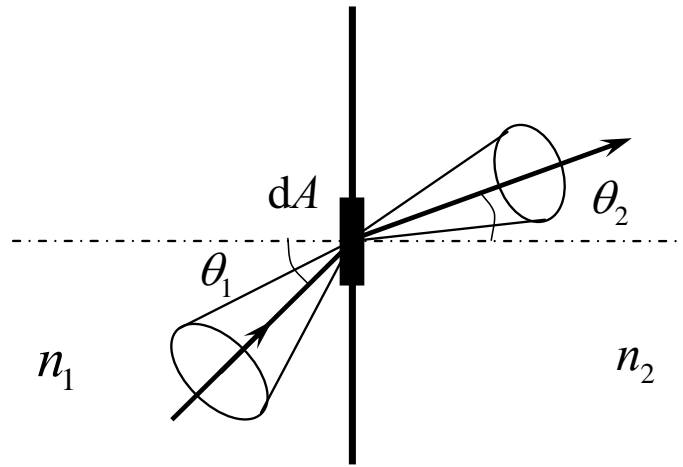

**Fig. 3.** Schematic of the two layer media to derive the relation between the energy flux and the transport intensity.

**Authors: Zhao, Tan and Liu**

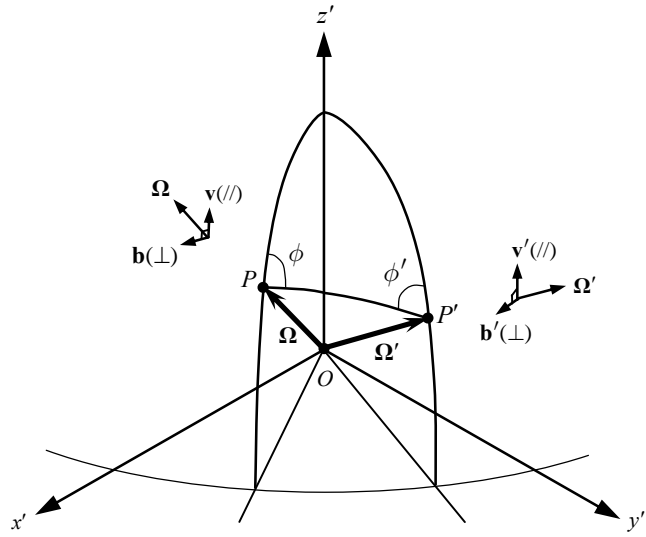

**Fig. 4.** Definitions of the Cartesian coordinate system.

**Authors: Zhao, Tan and Liu**

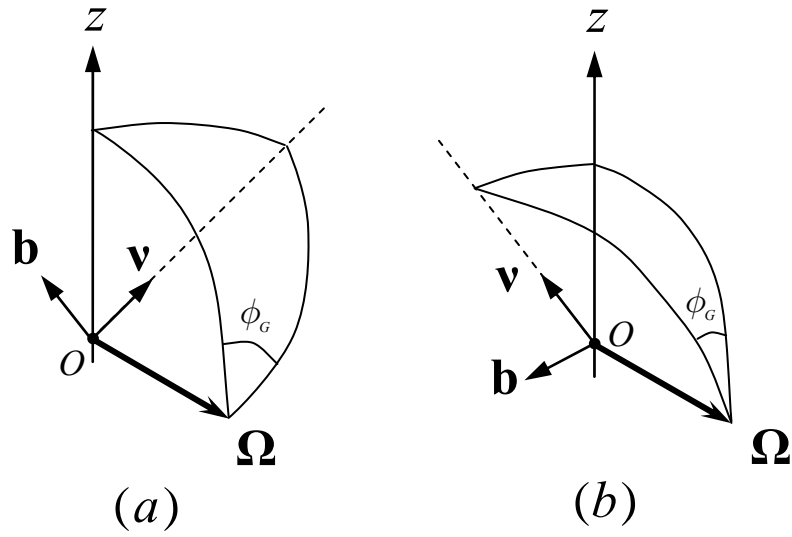

**Fig. 5.** Schematic to define a transformation of Stokes vector from local Frenet-Serret frame to a fixed frame. **(a)** Case 1, $(\mathbf{\Omega}\times\mathbf{v})\cdot\mathbf{e}_z > 0$, **(b)** Case 2, $(\mathbf{\Omega}\times\mathbf{v})\cdot\mathbf{e}_z < 0$.

**Authors: Zhao, Tan and Liu**

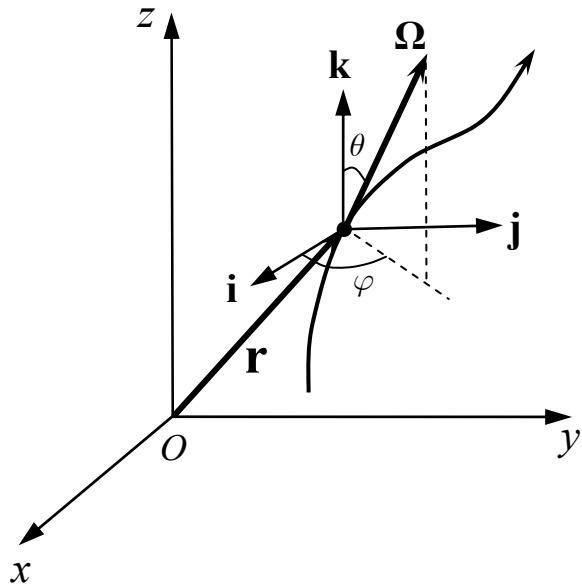

**Fig. 6.** Definitions of the Cartesian coordinate system.

**Authors: Zhao, Tan and Liu**

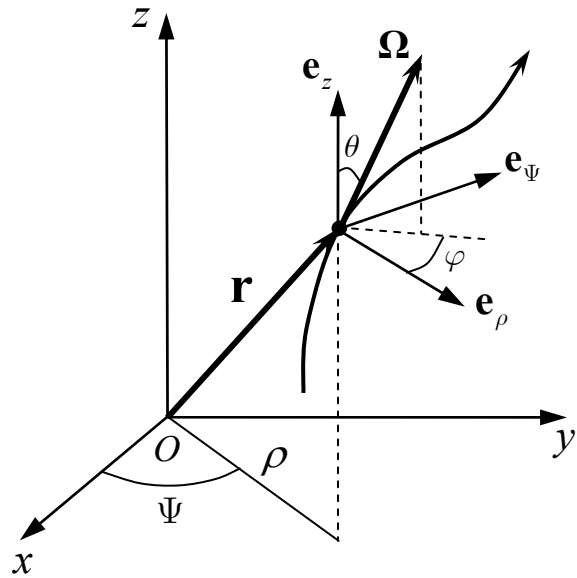

**Fig. 7(a).** Definitions of the traditional cylindrical coordinate system.

**Authors: Zhao, Tan and Liu**

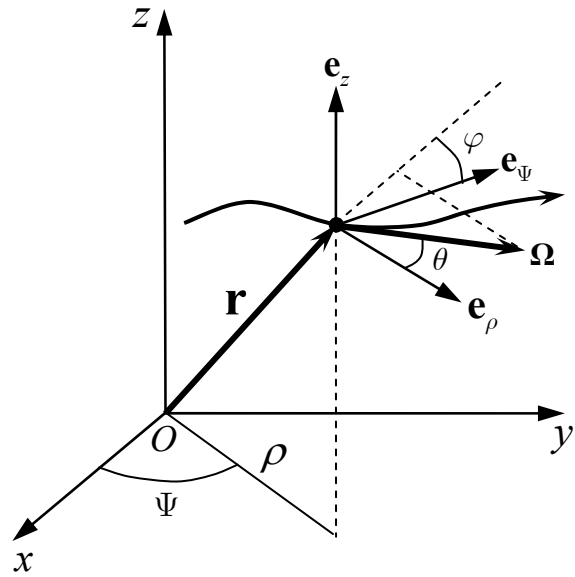

**Fig. 7(b).** Definitions of the new cylindrical coordinate system.

**Authors: Zhao, Tan and Liu**

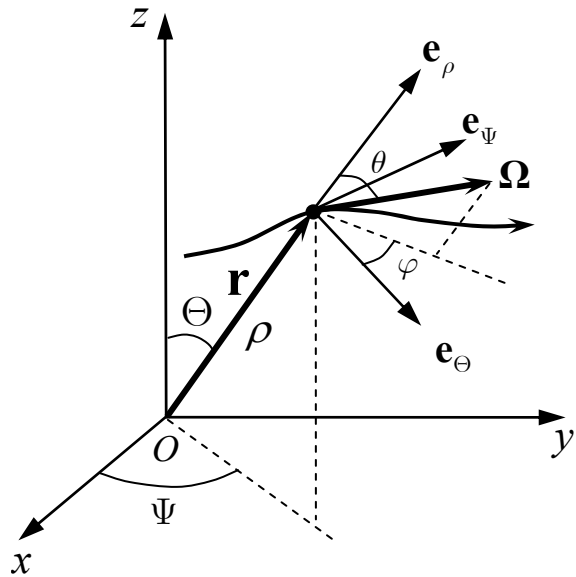

**Fig. 8.** Definitions of the spherical coordinate system.

**Authors: Zhao, Tan and Liu**

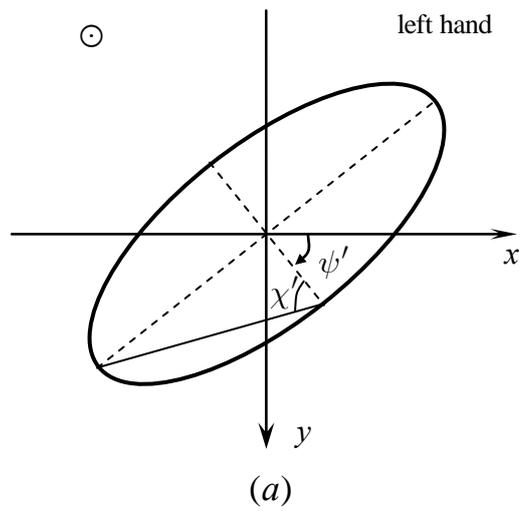

**Fig. 9(a).** Definitions of variables for the polarization ellipse in the left-hand frame configuration based on the short axis.

**Authors: Zhao, Tan and Liu**

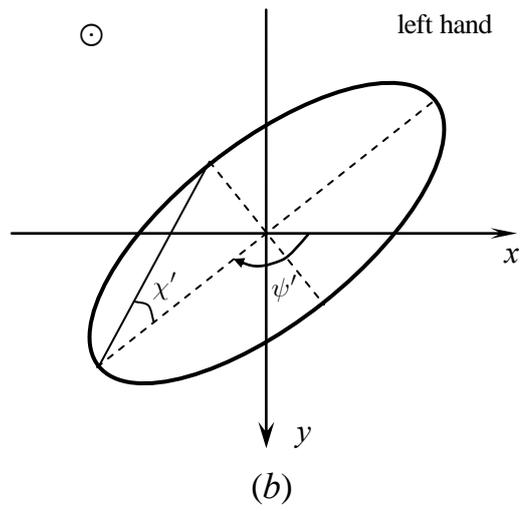

**Fig. 9(b).** Definitions of variables for the polarization ellipse in the left-hand frame configuration based on the long axis.

**Authors: Zhao, Tan and Liu**